\documentstyle[12pt]{article}
\newcommand{\be}{\begin{equation}}
\newcommand{\ee}{\end{equation}}
\begin{document}
\def\theequation{\arabic{section}.\arabic{equation}}
\begin{titlepage}
\title{Superquintessence}
\author{Valerio Faraoni$^{1,2}$ \\ \\
{\small \it $^1$ Physics Department, University of Northern British 
Columbia} \\ 
{\small \it 3333 University Way, Prince George, BC, V2N~4Z9, Canada} \\ \\
{\small \it $^2$ INFN-Laboratori Nazionali di Frascati, 
P.O. Box 13, I-00044 Frascati, Roma, Italy}
}
\date{} 
\maketitle   \thispagestyle{empty}  \vspace*{1truecm}
\begin{abstract} 
There is marginal evidence that the quintessential form of matter 
responsible for the acceleration of the universe observed today has ratio
between pressure and energy density $w<-1$. Such a regime, called 
superacceleration, cannot be achieved with conventional scalar field 
models. The simplest non-exotic model achieving superacceleration is that 
of a scalar field nonminimally coupled to the Ricci curvature. This model 
is studied for general potentials and an exact superaccelerating solution is 
presented. In quintessential inflation, the model can have blue 
gravitational wave spectra, improving the prospects for the detection of 
cosmological gravitational waves.
\end{abstract}
\vspace*{1truecm} 
\end{titlepage}   

\section{Introduction}
\setcounter{equation}{0}

Observations of the magnitude-redshift relation of distant Type Ia 
supernovae \cite{SN} led to the recent discovery that the universe 
undergoes accelerated  
expansion at the present era.
This acceleration regime is attributed to the influence of a yet unknown 
form of non-luminous matter called {\em quintessence}, which must have 
negative pressure \cite{quintessence}.  

Several theoretical models of quintessence have been proposed, including 
a cosmological constant, a time-varying energy density and, more 
plausibly, a scalar field \cite{quintessence}.

There is no shortage of scalar fields in modern particle physics and in 
gravitational theories; it suffices to think of the dilaton of string 
theories \cite{GreenSchwarzWitten}, of the supersymmetric partner of spin 
$1/2$ particles in supergravity, of Nambu-Goldstone bosons, of the Higgs 
field of the standard model, or of the inflaton field dominating the 
cosmic dynamics at early times \cite{LiddleLyth}, the scalar of geometric 
origin of Kaluza-Klein theories \cite{OverduinWesson}, or the Brans-Dicke 
scalar \cite{BD}. It is not surprising, therefore, that scalar fields 
have been widely employed as natural models of quintessence.

An accelerated expansion of the universe is obtained when the pressure 
$P$ and the energy density $\rho$ of the scalar field satisfy $w\equiv 
P/\rho <-1/3$. On the observational side, Ref.~\cite{Caldwell} pointed 
out that there is marginal evidence for values of the $w$-parameter less 
than $-1$. This range of values corresponds to a superexponential  
expansion that we call {\em superacceleration}.
An ordinary scalar field minimally coupled to the spacetime curvature, 
used in most models, cannot achieve such a regime (this will be shown in 
detail in Sec.~2); 
Ref.~\cite{Caldwell} therefore spurred interest in non-conventional, 
supergravity-inspired, scalar field models of quintessence exhibiting the 
``wrong'' sign of the kinetic energy density 
\cite{Chibaetal,Ziaeepour,White,RiazuloUzan}. 

In this paper it is demonstrated that a superacceleration regime can be 
achieved in a much simpler model containing only a scalar field nonminimally 
coupled 
to the Ricci curvature and with positive definite kinetic energy density. 
Such a model is not only esthetically appealing 
by virtue of simplicity, but is also rather compelling from a theoretical 
point of view, since nonminimal coupling necessarily arises due to 
first loop corrections (see Ref.~\cite{mypreprint} for  a recent review 
and references), or as a prediction of specific scalar field theories 
\cite{rengroup,FaraoniPRD96}; and even at the classical level, 
nonminimal couplig is required by the Einstein equivalence principle in 
general relativity \cite{SonegoFaraoni}.

The dynamics of a scalar field nonminimally coupled to gravity were 
recently investigated in a general dynamical system approach 
\cite{Amendolaetal,GunzigetalPRD,Peyresq,GunzigetalCQG,Tarcisioetal} and 
in the context of inflation 
\cite{FaraoniPLA2000,mypreprint,FaraoniPRD2000}. Several results 
scattered in these references are relevant for quintessence and, in 
particular, for the superacceleration regime. In the present work, the 
relevant aspects of this approach for superacceleration are collected and 
discussed, while  
new features are pointed out. Instead of assuming a specific form of the 
scalar field potential $V( \phi)$, as done in previous models of 
quintessence based on nonminimally coupled scalars 
\cite{PerrottaBaccigalupiMatarrese}, we keep the form of $V( \phi)$ 
completely general.

Theoretical evidence for superacceleration in nonminimally coupled models is 
pointed 
out, and an exact solution is derived which is superaccelerating and has 
an effective equation of state described by a time-dependent ratio 
$w(t)=P/\rho$.

The implications of present-day superacceleration (if confirmed) for 
quintessential 
inflationary scenarios of the universe are pointed out. Contrarily to 
minimally coupled scalar field models, blue gravitational wave spectra are 
possible; the resulting higher power at small scales makes nonminimally 
coupled scalar field models interesting from the point of view of the 
detection of primordial gravitational waves with present and future laser 
interferometers.

Certain features of the dynamics of nonminimally coupled scalars also 
occur in string-inspired pre-big bang cosmology \cite{Gasperinireview}. 
The latter was 
criticized on the basis of being unable to provide a true inflationary 
regime~-~the 
ratio of any physical length to the Planck length scale, that is  a true 
measure of inflation, actually decreases in pre-big bang cosmology, thus 
failing to solve the horizon problem \cite{Coule}. A similar problem is 
present in any dilaton-driven regime of cosmic acceleration, but is 
absent in the theory presented in this paper.

The plan of the paper is as follows: in Sec.~2 the effective equation of 
state of  a universe dominated by a scalar field is discussed, and it is 
shown that superacceleration is absent in a minimally coupled scalar 
field model. The simplest model of superquintessence (the material source 
fueling the yet unconfirmed, present-day, superacceleration regime of the 
universe) is then introduced, and the relevant equations are derived. In 
Sec.~3, an exact superaccelerating solution is presented: Sec.~4 
discusses the implications for quintessential inflationary models,  
and the gravitational wave spectra constituting a distinctive feature of 
superaccelerated cosmic expansion, and contains outlooks on nonminimally 
coupled scalar field models of quintessence.

\section{Superquintessence and superacceleration}
\setcounter{equation}{0}

It was  pointed out in Ref.~\cite{Caldwell} that models with $ w\equiv 
P/\rho 
< -1 $ are in agreement with  observations of Type Ia supernovae \cite{SN}, 
even for very negative $w$, 
and  there is marginal evidence that the pressure to density ratio is 
indeed less than $-1$. Certainly, the $w<-1$ parameter space is not 
excluded by the available observational data \cite{SN}. The possibility that 
$w<-1$ was investigated by several 
authors \cite{Caldwell,Chibaetal,Ziaeepour,White,RiazuloUzan} mainly using 
models based
on a scalar field with the ``wrong'' sign of the kinetic energy term in 
the Lagrangian. Dark energy density with the property $w<-1$ was called 
{\em phantom energy} by these authors.

The search for a rather exotic model of phantom energy is justified by 
the following considerations: $P<-\rho$ is equivalent to $\dot{H}>0$, 
where $H$ is the Hubble parameter and an overdot denotes differentiation 
with respect to the cosmic time, and we refer to a spatially flat 
Friedmann-Lemaitre-Robertson-Walker (FLRW)  
universe strongly favoured by cosmological observations. The spacetime 
metric is \cite{footnote1} 
\be  \label{metric}
ds^2=-dt^2+a^2(t) \left(  dx^2+dy^2+dz^2 \right) 
\ee
in comoving coordinates $\left( t,x,y,z\right)$, 
and $ H=\dot{a}/a $ satisfies the equations \cite{LiddleLyth} 
\be  \label{2}
H^2=\frac{\kappa}{3} \, \rho \;,
\ee
\be  \label{3}
\frac{\ddot{a}}{a}=\dot{H}+H^2=-\, \frac{\kappa}{6} \, \left( \rho +3P 
\right) \;, 
\ee
where $\kappa \equiv 8\pi G$ and $\rho$ 
and $P$ are, respectively, the energy density and pressure of the cosmic 
fluid. Eqs.~(\ref{2}) and (\ref{3}) imply that 
\be  \label{4}
\dot{H}=-\frac{\kappa}{2} \left( \rho + P \right)
\ee
and therefore 
\be
P<-\rho \; \Leftrightarrow w< -1 \; \Leftrightarrow \; \dot{H}>0 \;.
\ee
A regime with $\dot{H}>0$ was originally investigated in the context of 
inflationary models of the early universe and called {\em superinflation} 
\cite{LucchinMatarrese}, a name later adopted in string cosmology. Perhaps a 
better name would be {\em superacceleration}, to denote the possibility 
that $\dot{H}>0$ in a quintessence-dominated universe today, well after 
the end of inflation; we use this terminology throughout the present 
paper.

The simplest models of quintessence and inflation are based on a scalar 
field minimally coupled to gravity and are not adequate to describe a 
superacceleration regime; in fact, in such models, the scalar field 
$\phi(t)$ behaves as a perfect fluid with  
energy density and pressure 
\be \label{5}
\rho=\frac{\dot{\phi}^2}{2} +V( \phi ) \;,
\ee
\be \label{6}
P=\frac{\dot{\phi}^2}{2} - V( \phi ) \;,
\ee
respectively, where $V( \phi)$ is the scalar field potential. Eq.~(\ref{4}) 
becomes, in this case, 
\be  \label{7}
\dot{H}=-\, \frac{\kappa}{2}\, \dot{\phi}^2 \;.
\ee
Therefore, $\dot{H} \leq 0 $ in such models, with $\dot{H}=0$ 
corresponding to a de Sitter solution with scale factor $a(t)=a_0 \,  
\mbox{e}^{H t}$. The effective equation of state of a FLRW universe 
dominated by a minimally coupled scalar field is given, for a {\em 
general} potential $V( \phi)$, by
\be   \label{8}
\frac{P}{\rho}=\frac{\dot{\phi}^2-2V}{\dot{\phi}^2 +2V} \equiv w(x) \;,
\ee
where $x \equiv \dot{\phi}^2/2V$ is the ratio between the kinetic and the 
potential energy densities of the scalar $\phi$. Under the usual 
assumption $V \geq 0$ guaranteeing that the energy density of $\phi$ is 
non-negative, the function
\be  \label{9}
w(x)=\frac{x-1}{x+1} 
\ee
monotonically increases from its minimum $w_{min}=-1$ attained at $x=0$ 
to the horizontal asymptote $+1$ as $x\rightarrow +\infty$ (corresponding 
to $V=0$). The 
slow-rollover regime of inflationary models corresponds to $x \ll 1$ and to  
$w(x) $ near its minimum, where the kinetic energy density 
$\dot{\phi}^2/2 $ of $\phi$ is negligible in comparison to its potential 
energy density $V( \phi)$. As the kinetic energy density increases, the 
equation of state progressively deviates from $P=-\rho$ and the pressure 
becomes less and less negative. The system gradually moves away from the 
slow-rollover regime. At equipartition between the kinetic and the 
potential energy densities ($x=1$), one has the dust equation of state 
$P=0$. The pressure then  becomes positive as $x$ increases and, when the 
kinetic energy density completely dominates the potential energy density 
($x \ll 1$), one finally reaches the stiff equation of state $P=\rho$. 
Thus, for minimally coupled scalar fields, one encompasses the range 
of equations of state 
\be \label{10}
-1 \leq w \leq 1 
\ee
and observational data producing $w<-1 $ are not explained by this 
``canonical'' scalar field model, unless one is willing to accept a 
negative potential $V( \phi) < -\dot{\phi}^2/2 $,  which yields the 
negative energy density $\rho=\dot{\phi}^2/2 +V( \phi)$. This  
violation of the weak energy condition is very unappealing, and the  
possibility that $V<0$ is excluded from the rest of this work.

The superacceleration regimes studied in the literature consist of 
pole-like inflation with scale factor
\be  \label{11}
a(t)=\frac{a_0}{t-t_0} \;,
\ee
a  special form of superacceleration considered in early inflationary 
theories \cite{LucchinMatarrese}, in pre-big bang cosmology 
\cite{Gasperinireview}, and in Brans-Dicke theory \cite{Coule}.

Situations with $P<-\rho$ can also occur in higher derivative theories of 
gravity \cite{Pollock} or due to semiclassical particle production 
resulting in nonzero bulk viscosity \cite{Barrow}. In the words of
Ref.~\cite{Caldwell}, the marginal evidence for phantom energy with 
$ w <-1$ (which we prefer to call {\em superquintessence}, consistently 
with the word ``superacceleration'') poses the challenge of building a 
microphysical model of phantom energy (superquintessence). Rather 
unconventional, supergravity-inspired models with kinetic energy density 
of the scalar field $-\dot{\phi}^2/2 $ instead of $\dot{\phi}^2/2$ were 
investigated (\cite{Caldwell,Chibaetal,Ziaeepour,White}-see also 
\cite{ParkerRaval}).  
When the scalar field potential $V$ is absent, this form of matter is 
called {\em kinetically driven quintessence}. While there certainly is 
scope 
for investigating such models, Occam's razor dictates that  one should 
first study the simplest, most natural model of superquintessence; this 
is the very simple theory of a  scalar field nonminimally coupled  
to gravity, described by the action
\be  \label{12}
S=\int d^4x \sqrt{-g} \left[ \frac{R}{2\kappa} -\frac{1}{2} \nabla^c
\phi \nabla_c \phi -V( \phi ) -\frac{\xi}{2}R\phi^2 \right] \; , 
\ee
where $g$ is the determinant of the metric tensor and the kinetic 
energy density term is canonical, hence the latter is positive definite in a 
FLRW space. We do  not include 
other forms of matter in the action, as it would be necessary to build a 
complete model of quintessence. We assume that the quintessential field 
$\phi$ has already begun to dominate the dynamics of the universe. That 
this scenario is plausible, and that the cosmic coincidence problem can 
be solved in this context, was already shown in previous works 
\cite{PerrottaBaccigalupiMatarrese}.  Variation of 
the action with respect to $g_{ab} $ leads to the field equations
\be \label{13}
G_{ab}=\kappa T_{ab} \;,
\ee
where the scalar field energy-momentum tensor is 
\be \label{14}
T_{ab}=\nabla_a \phi \, \nabla_b \phi -\frac{1}{2} \, g_{ab} \nabla^c 
\phi \, \nabla_c \phi -V \, g_{ab} +\xi \left( g_{ab} \Box -\nabla_a 
\nabla_b \right) \left( \phi^2 \right) +\xi G_{ab} \phi^2 \; .
\ee
Note that, in the presence of nonminimal coupling (hereafter referred 
to as ``NMC''), there are three possible inequivalent  ways of writing the 
field equations (see Refs.~\cite{FaraoniPRD2000,BellucciFaraoni} for a 
recent discussion) 
and the scalar field stress-energy tensor. We choose the procedure 
described by eqs.~(\ref{13}) and (\ref{14}) because\\
{\em i)} the corresponding energy density
\be \label{15}
\rho= \frac{\dot{\phi}^2}{2} + V(  \phi)
+3\xi H\phi \left( H\phi+ 2 \dot{\phi}\right)
\ee
is always positive definite, since it is related to the Hubble parameter 
by the Hamiltonian constraint
\be  \label{16}
H^2=\frac{\kappa}{3} \, \rho \;,
\ee
which guarantees that $\rho \geq 0$. This is not the case with other 
definitions of energy density used in the literature, which has led to 
debate the validity of the weak energy condition 
\cite{BarceloVisserCQG,BellucciFaraoni,FordRomanpreprint} for nonminimally coupled 
classical scalars.\\\\
{\em ii)} The gravitational coupling is constant in the approach of this 
paper, while the effective gravitational coupling
\be \label{17}
\kappa_{eff}=\frac{\kappa}{1-\kappa\xi \phi^2}
\ee
used in other approaches and widely present in the literature can change 
sign and diverge, leading to spurious 
effects and loss of generality in the field equations when the scalar 
field attains the critical values
\be  \label{18}
\pm \phi_c=\pm \frac{1}{\sqrt{\kappa\xi}}
\ee
for $\xi >0$ (for examples see 
Refs.~\cite{GunzigetalCQG,BarceloVisserCQG}). The nonminimally coupled 
scalar field has pressure given by 
\be  \label{18bis} 
P=\frac{\dot{\phi}^2}{2}-V( \phi ) -\xi \left[ 4H\phi \dot{\phi} 
+2\dot{\phi}^2 +2\phi \ddot{\phi} +\left( 2\dot{H}+3H^2 \right) \phi^2 
\right] 
\ee 
and obeys the equations
\be  \label{fe1}
6\left[ 1 -\xi \left( 1- 6\xi \right) \kappa \phi^2
\right] \left( \dot{H} +2H^2 \right) 
-\kappa \left( 6\xi -1 \right) \dot{\phi}^2   
- 4 \kappa V  + 6\kappa \xi \phi V' = 0 \; ,
\ee
\begin{equation}  \label{fe2}
\frac{\kappa}{2}\,\dot{\phi}^2 + 6\xi\kappa H\phi\dot{\phi}
- 3H^2 \left( 1-\kappa \xi \phi^2 \right) + \kappa  V =0 \, ,
\end{equation}
\be  \label{KG}
\ddot{\phi}+3H\dot{\phi}+\xi R \phi +V' =0 \; .
\ee
The fact that a nonminimally coupled scalar field can generate a 
superacceleration regime was already known to the authors of 
Ref.~\cite{LucchinMatarrese} in the context of inflation (specifically, 
pole-like inflation), and was 
recently revisited in Refs.~\cite{GunzigetalPRD,Peyresq}; a 
dynamical systems approach clearly showed the possibility of 
superacceleration and numerical solutions with $\dot{H}>0$ were 
presented. It was speculated that such a regime could be important for 
the possible genesis of the universe from Minkowski space and for the 
modelling of 
quintessence. Further, in Ref.~\cite{Tarcisioetal}, an exact solution 
corresponding to integrability of the field equations and to the equation 
of state $P=-5\rho /3$ was found.

The fundamental differences between our discussion and previous ones are 
that we do not impose that the ratio $w=P/\rho$ be constant, and we do 
not choose a specific form of $V( \phi)$, but rather keep the discussion 
valid for {\em any} potential. Quintessence scenarios with a specific 
potential and NMC, as well as generalized Brans-Dicke couplings, or 
induced gravity, can be found in the literature, although 
superacceleration was not studied there 
\cite{PerrottaBaccigalupiMatarrese,BertolamiMartins}.

\section{An exact superaccelerating solution}
\setcounter{equation}{0}

The 
assumption $w=$const. used in the literature is useful to find analytical 
solutions, but is very restrictive: by imposing $w=$const. in a spatially 
flat FLRW universe one can only obtain power-law ($a=a_0t^p$) and de 
Sitter solutions (this is true for both minimal and nonminimal coupling), 
or  a third exact solution presented in Ref.~\cite{FaraoniAmJP}. Solutions 
for non-spatially flat universes and arbitrary values of $w$ are 
summarized in \cite{AmJP1}. The  search for analytical solutions 
probably constitutes the reason why only pole-like inflation (an inverse 
power-law) was discussed in the literature in conjunction with 
superacceleration \cite{LucchinMatarrese,Coule,Gasperinireview}. 
Superaccelerating solutions are more easily found numerically 
\cite{GunzigetalPRD,Peyresq}.

A new explicit solution of the 
field equations (\ref{fe1})-(\ref{KG}) is obtained by setting $\phi$ 
equal to one of the critical values $\pm \phi_c$ given by eq.~(\ref{18}); 
then the trace of the field equations
\be  \label{19}
R=6\left( \dot{H}+H^2 \right)=\kappa \left( \rho -3P \right)
\ee
yields
\be\label{20}
\dot{H}+2H^2=C \;,
\ee
where $C$ is a constant. For $C>0$ one has
\be \label{21}
\frac{\dot{H}}{1-2H^2/C}=C 
\ee
which is reduced to a quadrature as 
\be  \label{22}
\int\frac{dx}{1-x^2}=\sqrt{2C}\left( t-t_0 \right)  \,,
\ee
where $x\equiv \sqrt{2/C}\, H$. One has
\be   \label{23}
\mbox{arctanh} \, x= \sqrt{2C}\, \left(t-t_0 \right)
\ee
if $x^2>1$ and
\be   \label{24}
\mbox{arctanh} \, x= \ln \left[ \sqrt{ \frac{1+x}{1-x} } \, \right]  
\ee
if $x^2 <1$, and therefore
\be   \label{25}
H= \sqrt{\frac{C}{2}}\, \mbox{tanh} \left[ \sqrt{2C} \,  
\left(t-t_0 \right) \right] 
\ee
in both cases $H>\sqrt{C/2} $ and $H<\sqrt{C/2} $ (the cases $ H 
=\pm \sqrt{C/2}$ correspond to trivial de Sitter solutions). 
Eq.~(\ref{25}) contradicts the limit 
$H>\sqrt{C/2} $ and hence (\ref{25}) is only a solution for $H<\sqrt{C/2} 
$, corresponding to the scale factor
\be \label{26}
a=a_0 \cosh^{1/2} \left[ \sqrt{2C} \, \left( t-t_0 \right) \right] \;.
\ee
This solution describes an asymptotic contracting de Sitter space as $ 
t\longrightarrow -\infty$, reaching  a minimum size at 
$t=t_0$, and then expanding and superaccelerating with $\dot{H}>0$ and 
$w<-1$. As $t  \longrightarrow+\infty$, the solution approaches an 
expanding de Sitter space. The effective equation of state is 
time-dependent, with
\be  \label{27}
w(t)=\frac{P}{\rho}=-\, \frac{A+\xi \left( 2\dot{H}+3H^2 \right)}{3H^2+A}=
C\left\{ 2-\frac{1}{2} \tanh^2 \left[ \sqrt{2C}\, \left( t-t_0 \right) 
\right] \right\} \;,
\ee
where $A=\kappa V( \pm \phi_c )$.  While this exact solution is clearly 
fine-tuned in the scalar field value, it has the merit of explicitly 
illustrating the superacceleration phenomenon; the latter is a generic 
feature of the dynamics of  a nonminimally coupled scalar field. This was 
shown by considering the natural potential
\be  \label{28}
V( \phi )=\frac{m^2}{2} \, \phi^2 + \lambda \phi^4
\ee
and the conformal case $\xi=1/6$ (the value of $\xi$ that is an 
infrared fixed point of the renormalization group \cite{Ford,rengroup} 
and is required by the Einstein equivalence principle \cite{SonegoFaraoni}).
The dynamics of this system are much richer than in the minimally coupled 
($\xi=0$) case and were studied in detail 
\cite{GunzigetalCQG,GunzigetalPRD,Peyresq,Amendolaetal}.

\section{Outlooks}
\setcounter{equation}{0}

If the nonminimally coupled quintessence field was also responsible 
for inflation early in the history of the universe, as in certain 
quintessential inflationary scenarios \cite{PeeblesVilenkin}, one has to 
verify 
that the slow-roll approximation necessary to build a general theory of 
inflation is meaningful. In other words, one should check that the de Sitter 
solutions behaving as  attractor points in phase space when $\xi=0$, remain 
attractors when NMC is added to the theory. This is true subject 
to certain constraints on the value of the coupling constant $\xi$ and the 
potential $V( \phi) $ \cite{FaraoniPLA2000,mypreprint}. The following 
dimensionless slow-roll parameters can be introduced \cite{Hwang,mypreprint}
\be  \label{29}
\epsilon_1 = \frac{\dot{H}}{H^2}\; , 
\ee
\be  \label{30}
\epsilon_2 =\frac{\ddot{\phi}}{H \dot{\phi}}  \; , 
\ee
\be  \label{31}
\epsilon_3 =-\, \frac{\xi \kappa \phi \dot{\phi}}{H \left[ 1- \left(
\frac{\phi}{\phi_1} \right)^2 \right]} \; , 
\ee
\be  \label{32}
\epsilon_4 =-\, \frac{\xi \left( 1-6\xi \right) \kappa \phi \dot{\phi}}
{H \left[ 1-\left( \frac{\phi}{\phi_2} \right)^2 \right] } \; ,
\ee
where 
\be  \label{33}
\pm \phi_2 \equiv \frac{\pm1}{\sqrt{ \kappa\xi \left( 1-6\xi \right)}} 
\ee
for $0<\xi<1/6$.
$\epsilon_3$ and $\epsilon_4$  vanish in the limit $\xi \rightarrow 0 $ of  
ordinary inflation; $\epsilon_4$
also  vanishes for conformal coupling ($\xi=1/6$). One has 
$\left| \epsilon_i \right|
<< 1$ ($i=1,2,3,4$) for every solution attracted by expanding de
Sitter spaces $\left( H_0, \phi_0 \right)$ for suitable values of $\xi $ 
and $V$ \cite{FaraoniPLA2000,mypreprint}. 
Moreover, $\epsilon_i=0$ exactly for de Sitter solutions with constant 
scalar field \cite{footnote2}. 

The spectral indices of scalar and tensor perturbations \cite{LiddleLyth} 
are expressed in terms of the slow-roll parameters 
\cite{Hwang,mypreprint} by
\be \label{34}
n_S=1+2\left( 2\epsilon_1-\epsilon_2+\epsilon_3-\epsilon_4 \right) \;,
\ee
\be \label{35}
n_T=2\left( 2\epsilon_1 -\epsilon_3 \right) \;.
\ee
In a superacceleration regime during inflation, with $\dot{H}>0$, one 
has $\epsilon_1>0$ 
and the spectral index of gravitational waves $n_T$ given by 
eq.~(\ref{35}) can be positive ({\em blue spectrum}), with more power at 
small wavelenghts than the usual inflationary perturbation spectra. This 
feature is very attractive for the gravitational wave community because 
it enhances the possibility of detection of cosmological gravitational 
waves. Detailed calculations for specific potentials (beyond the purpose 
of this work) are required in order to assess the resulting 
gravitational 
wave amplitudes in the frequency bands covered by the {\em LIGO}, {\em 
VIRGO}, and the other  present 
laser interferometry experiments; it seems that blue spectra are  
meaningful at 
least for  the future space-based interferometers \cite{gw}. Blue spectra 
are impossible with minimal coupling \cite{footnote3}, for which one obtains
$n_T=4\dot{H}/H^2  \leq 0$.

Another interesting feature emerging from the studies of 
Refs.~\cite{GunzigetalPRD,Peyresq} is the {\em spontaneous} 
exit of the universe from the superacceleration regime, to enter an 
ordinary accelerated one, followed again by spontaneous exit and entry 
into a 
decelerated epoch. These features are absent in string-based models that 
keep accelerating forever \cite{Fischeretal,Hellermanetal}, or in 
inflationary models where inflation is terminated by an {\em ad 
hoc}  modification of the equations or of the shape of $V( \phi) $.

Finally, we remark that the $\dot{H}>0$ regime achieved with NMC is  
a true superacceleration regime, contrarily to the case of pre-big bang 
cosmology and of Brans-Dicke theory. As pointed out in Ref. \cite{Coule}  
in these theories, while the universe expands, the Planck length also 
grows and the ratio of any physical length to the Planck length (which 
is a true measure of the inflation of the universe) actually {\em 
decreases}. In string-inspired pre-big bang cosmology this phenomenon is 
due to the fact that the gravitational coupling is $G\mbox{e}^{\Phi}$, 
where $\Phi (t) $ is the string dilaton \cite{Coule}.

For a nonminimally coupled scalar field, using the approach selected here 
and the corresponding $T_{ab}$ for the scalar field, the gravitational 
coupling is constant and there is no variation of the Planck length; the 
superacceleration regime is a true one.
A different approach using the effective gravitational coupling 
$ G_{eff}=G \left( 1-8\pi G \xi \phi^2 \right)^{-2}$  common 
in the literature, would suffer of the same problem of pre-big bang 
cosmology for $\xi>0$ and could not produce the exact solution (\ref{26}) 
since $G_{eff}$ diverges for the corresponding scalar field values $\pm 
\phi_c$.

Were the inequality $w<-1$ supported by future observations of distant 
supernovae, the 
conventional quintessence models based on minimally coupled scalars would 
have to be abandoned in favour of alternative models: among these, the 
nonminimally coupled theory here described appears as the simplest and 
most promising.

\section*{Acknowledgments}

It is  a pleasure to thank S. Matarrese, C. Baccigalupi and D. Coule for 
stimulating discussions.


\clearpage

\begin{thebibliography}{99}

\bibitem{SN} S. Perlmutter {\em et al.}, {\em Nature} {\bf 391}, 51 
(1998); A.G. Riess {\em et al.}, {\em Astrophys. J.} {\bf 116}, 1009 
(1998); {\em Astron. J.} {\bf 116}, 1009 (1998); {\em Astrophys. J} {\bf 
118}, 2668 (1999); B.R. Schmidt {\em et al.}, {\em Astrophys. J.} {\bf 
507}, 46 (1998).

\bibitem{quintessence} I. Zlatev, L. Wang and P.J. Steinhardt, {\em Phys. 
Rev. Lett.} {\bf 82}, 896 (1999); P.J. Steinhardt, L. Wang and I. Zlatev, 
{\em Phys. Rev. D} {\bf 59}, 123504 (1999); S.M. Carroll, {\em Phys. Rev. 
Lett.} {\bf 81}, 3067 (1998); L. Wang and P.J. Steinhardt, {\em 
Astrophys. J.} {\bf 508}, 483 (1998); V. Sahni and L. Wang, preprint 
astro-ph/9910097.

\bibitem{GreenSchwarzWitten} B. Green, J.M. Schwarz and E. Witten, {\em 
Superstring Theory} (Cambridge University Press, Cambridge, 1987).

\bibitem{LiddleLyth} A.R. Liddle and D. Lyth, {\em Cosmological Inflation 
and Large Scale Structure} (Cambridge University Press, Cambridge, 2000); 
E.W. Kolb and M.S. Turner, {\em The Early Universe} (Addison-Wesley, MA, 
1994).

\bibitem{OverduinWesson} J.M. Overduin and P.S. Wesson, {\em Phys. Rep.} 
{\bf 283}, 303 (1997).

\bibitem{BD} C.H. Brans and R.H. Dicke, {\em Phys. Rev.} {\bf 124}, 925 
(1961).

 \bibitem{Caldwell} R.R. Caldwell, preprint astro-ph/9908168.

\bibitem{Chibaetal} T. Chiba, T. Okabe and M. Yamaguchi, {\em Phys. Rev. 
D} {\bf 62}, 023511 (2000).

\bibitem{Ziaeepour} H. Ziaeepour, preprint astro-ph/0002400.

\bibitem{White} E. Schultz and M. White, preprint astro-ph/0104112.

\bibitem{RiazuloUzan} A. Riazulo and J.-P. Uzan, {\em Phys. Rev. 
D} {\bf 62}, 083506 (2000).

\bibitem{mypreprint} V. Faraoni, preprint hep-th/0009053.

\bibitem{rengroup}  
I.L. Buchbinder and S.D. Odintsov, {\em Sov. J. Nucl. Phys.} 
{\bf 40}, 848 (1983); 
I.L. Buchbinder and S.D. Odintsov, {\em Lett. Nuovo
Cimento} {\bf 42}, 379 (1985); 
I.L. Buchbinder, {\em Fortschr. Phys.} {\bf 34}, 605 (1986); 
I.L. Buchbinder, S.D. Odintsov and I.L. Shapiro, in {\em
Group-Theoretical Methods in Physics}, M.
Markov ed. (Nauka, Moscow, 1986); 
S.D. Odintsov, {\em Fortschr. Phys.} {\bf 39}, 621 (1991); 
T.S. Muta and S.D. Odintsov, {\em Mod.
Phys. Lett. A} {\bf 6}, 3641 (1991);
E. Elizalde and S.D. Odintsov, {\em Phys.
Lett. B} {\bf 333}, 331 (1994);
I. Buchbinder, S.O. Odintsov and I. Shapiro, {\em
Effective Action in Quantum Gravity} (IOP Publishing, Bristol, 1992);
I. Buchbinder, S.O. Odintsov and I. Lichzier, {\em
Class. Quant. Grav.} {\bf 6}, 605 (1989).

\bibitem{FaraoniPRD96} V. Faraoni, {\em Phys. Rev. D} {\bf 53}, 6813 
(1996); in  {\em Proceedings of the 7th Canadian Conference on General 
Relativity and  Relativistic Astrophysics}, D. Hobill ed. (The 
University of Calgary Press, Calgary, 1998).

\bibitem{SonegoFaraoni} S. Sonego and V. Faraoni, {\em Class. Quant. 
Grav.} {\bf 10}, 1185 (1993); V. Faraoni and S. Sonego, in {\em 
Proceedings of the 5th Canadian Conference on General Relativity and 
Relativistic Astrophysics}, R.B. Mann and R.G. McLenaghan eds. (World 
Scientific, Singapore, 1994).

\bibitem{Amendolaetal} L. Amendola, M. Litterio and F. Occhionero, {\em 
Int. J. Mod. Phys. A} {\bf 5}, 3861 (1990).

\bibitem{GunzigetalPRD} E. Gunzig, A. Saa, L. Brenig, V. Faraoni, T.M. Rocha Filho
and A. Figueiredo, {\em Phys. Rev. D} {\bf 
63}, 067301  (2001).

\bibitem{Peyresq} A. Saa, E. Gunzig, L. Brenig, V. Faraoni, T.M. Rocha Filho and A.
Figueiredo, preprint gr-qc/0012085.

\bibitem{GunzigetalCQG} E. Gunzig, V. Faraoni, T.M. Rocha Filho, A. Figueiredo and
L. Brenig, {\em Class. Quant. Grav.} 
{\bf 17}, 1783 (2000).

\bibitem{Tarcisioetal} T.M. Rocha Filho, A. Figueiredo, L. Brenig, E. Gunzig and V. 
Faraoni, {\em Int. J. Theor. Phys.} {\bf 39}, 1933 (2000).

\bibitem{FaraoniPLA2000} V. Faraoni, {\em Phys. Lett. A} {\bf 269}, 
209 (2000).

\bibitem{FaraoniPRD2000} V. Faraoni, {\em Phys. Rev. D} {\bf 62}, 
023504 (2000).

\bibitem{PerrottaBaccigalupiMatarrese} F. Perrotta, C. Baccigalupi and 
S. Matarrese, {\em Phys. Rev. D} {\bf 61}, 023507 (1999); C. Baccigalupi, 
S. Matarrese and F. Perrotta, {\em Phys. Rev. D} {\bf 62}, 123510 (2000); 
J.-P. Uzan, {\em 
Phys. Rev. D} {\bf 59}, 123510 (1999); S. Sen and T.R. Seshadri, preprint 
gr-qc/0007079.

\bibitem{Gasperinireview} M. Gasperini and G. Veneziano, {\em Astropart. 
Phys.} {\bf 1}, 317 (1993); M. Gasperini, preprint gr-qc/9706037; S.J. 
Rey, preprint hep-th/9609115; J.J. Levin, preprint gr-qc/9506017; see 
also the papers in http://www.to.infn.it/teorici/gasperini

\bibitem{Coule} D.H. Coule, {\em Phys. Lett. 
B} {\bf 450}, 48 (1999).

\bibitem{footnote1} $G$ is Newton's constant and, apart from
minor differences, we adopt the
notations and conventions of Ref.~\cite{Wald}. The metric signature 
is --~+~+~+, the speed of light and
Planck's constant assume the value
unity and $m_{pl}=G^{-1/2} $ is the Planck mass.  The components of the
Ricci tensor are given in terms of the
Christoffel symbols
$\Gamma_{\alpha\beta}^{\delta}$ 
by $R_{\mu\rho}=
\Gamma^{\nu}_{\mu\rho ,\nu}-\Gamma^{\nu}_{\nu\rho ,\mu}+
\Gamma^{\alpha}_{\mu\rho}\Gamma^{\nu}_{\alpha\nu}-
\Gamma^{\alpha}_{\nu\rho}\Gamma^{\nu}_{\alpha\mu} $, and  
$\Box \equiv g^{ab}\nabla_{a}\nabla_{b}$.

\bibitem{Wald} R.M. Wald, {\em General Relativity} (The University of 
Chicago Press, Chicago, 1984).

\bibitem{LucchinMatarrese} F. Lucchin and S. Matarrese, {\em Phys. Lett. 
B} {\bf 164}, 82 (1985).

\bibitem{Pollock} M.D. Pollock, {\em Phys. Lett. B} {\bf 215}, 635 (1988).

\bibitem{Barrow} J.D. Barrow, {\em Nucl. Phys. B} {\bf 310}, 743 (1988).

\bibitem{ParkerRaval} L. Parker and A. Raval, {\em Phys. Rev. D} {\bf 
62}, 083503 (2000).

\bibitem{BellucciFaraoni} S. Bellucci and V. Faraoni, preprint 
hep-th/0106108.

\bibitem{BarceloVisserCQG} C. Barcelo and M. Visser, {\em Class. Quant. 
Grav.} {\bf 17}, 3843 (2000). 

\bibitem{FordRomanpreprint} L.H. Ford and T.A. Roman, preprint gr-qc/0009076.

\bibitem{BertolamiMartins} O. Bertolami and P.J. Martins, {\em Phys. Rev. 
D} {\bf 61}, 064007 (2001).

\bibitem{FaraoniAmJP} V. Faraoni, {\em Am. J. Phys.} {\bf 69}, 372 (2001).

\bibitem{AmJP1} V. Faraoni, {\em Am. J. Phys.} {\bf 67}, 732 (1999).

\bibitem{Ford} L.H. Ford and D.J. Toms, {\em Phys. Rev. D} {\bf 25}, 1510 
(1982); L. Parker and D.J. Toms, {\em Phys. Rev. D} {\bf 32}, 1409 (1985).

\bibitem{PeeblesVilenkin} P.J.E. Peebles and A. Vilenkin, {\em Phys. Rev. 
D} {\bf 59}, 063505 (1999).

\bibitem{Hwang} J.C. Hwang, {\em Class. Quant. Grav.} {\bf 7}, 1613 (1990);
{\em Phys. Rev. D} {\bf 42}, 2601 (1990); 
{\em Phys. Rev. D} {\bf 53}, 762 (1996); 
{\em Class. Quant. Grav.} {\bf 14}, 1981 (1997); 
{\em Class. Quant. Grav.} {\bf 14}, 3327 (1997); 
{\em Class. Quant. Grav.} {\bf 15}, 1401 (1998); 
{\em Class. Quant. Grav.} {\bf 15}, 1387 (1998); 
J. Hwang and H. Noh, {\em Phys. Rev. D} {\bf 54}, 1460 (1996); 
D.I. Kaiser, {\em Phys. Rev. D} {\bf 52}, 4295 (1995); preprint 
astro-ph/9507048.

\bibitem{footnote2} In the presence of NMC, contrarily to the
$\xi=0$ 
case, de Sitter solutions with non-constant scalar field $\left( H_0, 
\phi (t) \right) $ are possible, but they are not attractors.

\bibitem{gw} {\em Proceedings of the 3rd Edoardo Amaldi Conference}, S. 
Meshkov {\em et al.} eds. (Pasadena, USA 1999), in press; B. Caron {\em et 
al.}, in {\em 1st International LISA Symposium on Gravitational Waves}  
(Oxfordshire, England, 1996), {\em Class. Quant. Grav.} {\bf 14}, 1461 
(1997).

\bibitem{footnote3} The possibility of blue 
spectra in the context of pole-like inflation was already noted in 
Ref.~\cite{LucchinMatarrese}.

\bibitem{Fischeretal} W. Fischer {\em et al.}, preprint hep-th/0104181.

\bibitem{Hellermanetal} S. Hellerman, N. Kaloper and L. Susskind, 
preprint hep-th/0104180.


\end{thebibliography}
\end{document}